\documentclass[%
preprint,
 amsmath,amssymb,
aps,
prb,
showkeys,
endfloats*,
multicol,
]{revtex4-2}

\usepackage[dvipdfmx]{color}
\usepackage{graphicx}
\usepackage{dcolumn}
\usepackage{bm}
\usepackage{color}
\usepackage{txfonts}

\begin{document}
\title{Theoretical investigation of interface atomic structure of graphene on NiFe alloy substrate}
\author{Naohiro Matsumoto}
\email{246t257t@stu.kobe-u.ac.jp}
\affiliation{Department of Electrical and Electronic Engineering, Graduate School of Engineering, Kobe University, Nada, Kobe 657-8501, Japan}
\author{Ryusuke Endo}
\affiliation{Department of Electrical and Electronic Engineering, Graduate School of Engineering, Kobe University, Nada, Kobe 657-8501, Japan}
\author{Mitsuharu Uemoto}
\email{uemoto@eedept.kobe-u.ac.jp}
\affiliation{Department of Electrical and Electronic Engineering, Graduate School of Engineering, Kobe University, Nada, Kobe 657-8501, Japan}
\author{Tomoya Ono}
\email{t.ono@eedept.kobe-u.ac.jp}
\affiliation{Department of Electrical and Electronic Engineering, Graduate School of Engineering, Kobe University, Nada, Kobe 657-8501, Japan}

\begin{abstract}
The implementation of graphene as tunnel barriers in tunneling magnetoresistance devices attracts attention. So far, two processes have been proposed to fabricate graphene/NiFe alloy interfaces. One is the transfer of graphene and the other is the evaporation of alloys onto graphene. The formation energy of a NiFe alloy substrate and the adsorption energy of graphene on the NiFe alloy substrate are investigated by a density functional theory calculations to reveal the difference in the atomic structure of the interface between the two processes. In the case of bare substrate, Ni-rich surfaces are preferable regardless of composition of substrates. Interestingly, for the graphene adsorbed substrates, Fe-rich surfaces are stable due to the hybridization of $p_z$ orbital of C atom and $d_{z^2}$ orbital of Fe atom. This result indicates that the Ni-rich (Fe-rich) interface is formed by the graphene-transfer process (the alloy-evaporation process) and the composition ratio of the surface layer of graphene/NiFe alloy interfaces is significantly affected and controlled by the fabrication process.
\end{abstract}

\maketitle

\section{Introduction}
Tunneling magnetoresistance (TMR) is a significant phase in spintronics in magnetic tunnel junctions (MTJs),\cite{NatureMater_6_000813,JMagMagMater_321_000555,JPhysDApplPhys_46_074001} each of which consists of an ultrathin tunnel barrier separating two ferromagnetic (FM) metal electrodes with a variable magnetization direction. The key element of an MTJ is its tunnel barrier, and the most commonly used materials for the tunnel barrier remain to be MgO\cite{JMagMagMater_200_000248,JMagMagMater_126_000524,JApplPhys_101_09B501} or Al$_2$O$_3$.\cite{PhysRevB_63_054416,NatureMater_3_000862,NatureMater_3_000868,ApplPhysLett_93_082508,NatureMater_9_000721} Recently, two-dimensional (2D) materials, with atomically thin layers, have attracted considerable attention as useful nonmagnetic materials for tunnel barriers in MTJs.\cite{PhysRevB_90_041401,ApplPhysRev_08_021308,NatureNanotechnology_16_856,AdvTheorySiml_05_2200178,npj2DMaterApp_06_62,PhysRevLett_99_176602,PhysRevB_80_035408,IEEETransMagn_44_002624,NanoLett_12_003000,ACSNano_6_010930,IEEETransMagn_44_002624,NanoLett_12_003000,ACSNano_6_010930,JMagMagMater_484_000462,JpnJApplPhys_58_SDDE01,ApplSurfSci_510_145315,ElectronMaterLett_18_000313,ACSNano_16_014007,PhysicaB_655_414740,NanoResearch_8_001357,NanoscaleAdv_04_000117,JPhysD_58_165303}, Graphene\cite{PhysRevLett_99_176602,PhysRevB_80_035408,IEEETransMagn_44_002624,NanoLett_12_003000,ACSNano_6_010930,IEEETransMagn_44_002624,NanoLett_12_003000,ACSNano_6_010930,JMagMagMater_484_000462,JpnJApplPhys_58_SDDE01,ApplSurfSci_510_145315,ElectronMaterLett_18_000313,ACSNano_16_014007,PhysicaB_655_414740} and hexagonal boron nitride (hBN)\cite{PhysRevB_80_035408,NanoResearch_8_001357,NanoscaleAdv_04_000117,ElectronMaterLett_18_000313,JPhysD_58_165303} are used as the tunnel barrier in MTJs producing considerable TMR signals with good stability. To realize a high TMR ratio, it is expected that one should create contamination- and oxidation-free 2D material/FM metal electrode interface. One of the common procedures to fabricate the interface is the transfer process, in which exfoliated 2D materials are transferred to the FM metal substrates from other materials.\cite{NanoResearch_8_001357,JMaterChemC_4_008711} However, the contaminations and surface oxidation of the FM metal substrates occur during the transfer process. Indeed, any undesired surface oxidation of typical FM metals (Ni, Co, Fe, and their alloys) quenches their delicate spin polarization, rendering them useless for spintronics.\cite{JPhysSocJpn_77_031001} Asshoff {\it et al.} proposed the procedure where a 2D material flake is suspended over a SiN$_x$ membrane and then the Co and NiFe electrodes are evaporated onto the suspended flake from the top and bottom sides, respectively.\cite{2DMaterials_4_031004} Emoto {\it et al.} fabricated an MTJ by growing few-layer hBN on the FM electrodes and the Co thin film is evaporated as the top electrode.\cite{ACSApplMaterInterfaces_16_031457} Although the fabrication process is different, a NiFe alloy is used as the bottom electrode in both the procedures. NiFe alloy is widely used for free layers for MTJs owing its low coercivity. In addition, Emoto {\it et al.} used NiFe alloy films to form few-layer hBN because pure Fe electrodes provide relatively thick hBN multilayers. Although the spin polarization of the electrode surface plays an important role in realizing excellent spin coherency and high TMR ratio, the atomic and electronic structures of the 2D material/NiFe alloy interface are still not fully understood.

In this study, the density functional theory (DFT)\cite{PhysRev_136_B864} calculations are carried out to investigate the atomic and electronic structures of the graphene/NiFe alloy interface. It is found that the Ni-rich condition is preferable in the case of a bare substrate. On the other hand, a surface fully covered by Fe atoms is the most stable when graphene is adsorbed. C atoms of graphene sit on top and hcp sites of the (111) plane. The difference in the adsorption energy of the graphene/NiFe alloy interface can be explained by the occupation of bonding states using the partial density of states (PDOS). Our result implies that the fabrication process of the graphene/NiFe interfaces affects the composition ratio of the surface layer of FM metal as well as its magnetic property.

The rest of this paper is organized as follows. In Sec.~\ref{sec:Method}, the computational models and methods are explained. The interface atomic and electronic structures and adsorption energy are discussed in Sec.~\ref{sec:Results and discussion}. Finally, we summarize our findings in Sec.~\ref{sec:Conclusion}.

\section{Method}
\label{sec:Method}
\begin{figure}
\includegraphics{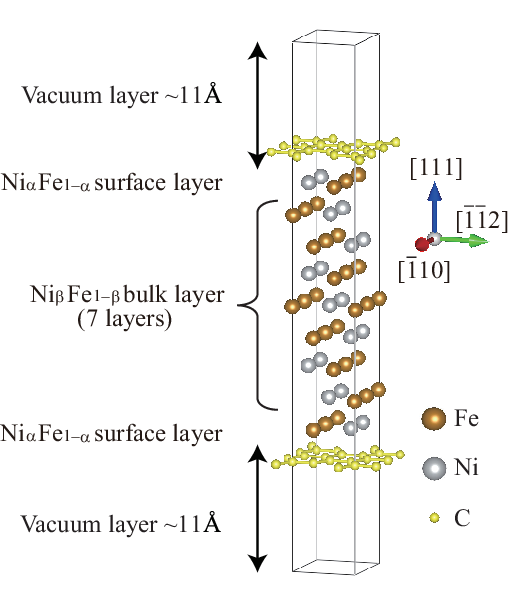}
\caption{Overall view of computational model. The case of $\alpha=0.50$ and $\beta=0.50$ is shown as an example.}
\label{fig:overview of model}
\end{figure}

\begin{figure}[htbp]
\includegraphics{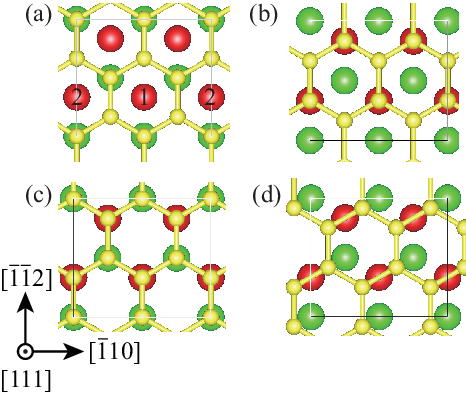}
\caption{Adsorption sites of graphene. (a) hcp-fcc site, (b) top-fcc site, (c) top-hcp site, and (d) bridge-top site. Red (green) spheres are atoms in the surface layer (topmost atoms of the bulk layer) of the NiFe alloy substrate and yellow spheres are C atoms in graphene. The atoms pointed by 1 and 2 (1) in (a) are Fe in the case of $\alpha$=0.50 ($\alpha$=0.75).}
\label{fig:Adsorption sites}
\end{figure}

Figure~\ref{fig:overview of model} shows the computational model of the graphene/NiFe alloy interface. The bare metal substrate consists of nine atomic layers, in which the topmost and bottommost layers are named the surface layers and the remaining layer is named as the bulk layer. The surface layers are a binary compound of Ni$_\alpha$Fe$_{1-\alpha}$ ($\alpha$=0.00, 0.25, 0.50, 0.75, and 1.00) and the bulk layer is a binary compound of Ni$_\beta$Fe$_{1-\beta}$ ($\beta$=0.50, 0.75, and 1.00). Fe atoms lie on a ($\bar{1}$10) plane in the bulk layer of Ni$_3$Fe and the atomic configuration of the bulk layer for NiFe is $L1_0$. The atomic configurations of Ni and Fe atoms in the surface layers are set so that the stacking sequence of Ni$_\beta$Fe$_{1-\beta}$ is kept when $\alpha=\beta$. As shown in supplementary data, we have assured that our conclusion is not affected when $L1_2$ structure is used for Ni$_3$Fe. Monolayer graphenes are adsorbed on both the edges of the substrate. Four possible adsorption geometries with C atoms located at the hcp-fcc, top-fcc, top-hcp, and bridge-top sites are considered as shown in Fig.~\ref{fig:Adsorption sites}. All calculations are performed using the DFT-based code Vienna Ab initio Simulation Package (VASP).\cite{PhysRevB_47_000558,PhysRevB_54_011169,PhysRevB_59_001758} Electronic exchange and correlation interactions are included through the Perdew--Burke--Ernzerhof generalized gradient approximation.\cite{PhysRevLett_77_003865} The electron--ion interactions are described by the projector-augmented wave method,\cite{PhysRevB_50_017953} and spin polarization is considered with a plane wave cutoff of 550 eV. The van der Waals dispersion interactions are treated using the DFT-D2 method of Grimme\cite{JComputChem_27_001787} as implemented in the VASP code. The supercell size is $\sqrt{2}a_b \times \sqrt{6}a_b/2 \times 36.34 \: $\AA$^3$, where $a_b$ (=3.524 \AA) is the experimental lattice constant of the Ni fcc bulk. The Brillouin-zone integration is sampled via a 19 $\times$ 19 $\times$ 1 Monkhorst--Pack mesh of $k$ points.\cite{PhysRevB_13_005188} We implement structural optimization for the atoms in the surface layers and graphenes until all the force components decrease to below 0.02 eV/\AA.

\section{Results and discussion}
\label{sec:Results and discussion}
\subsection{Formation energy of bare NiFe alloy substrate}
\label{subsec:Formation energy of bare NiFe alloy substrate}
\begin{figure}[htbp]
\includegraphics{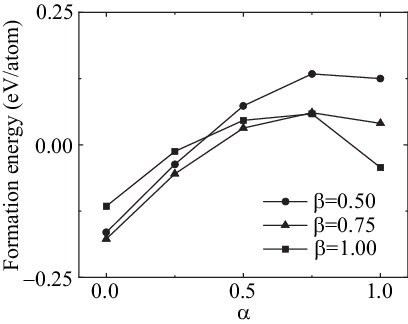}
\caption{Formation energy of bare NiFe alloy substrate.}
\label{fig:Total energy of NiFe}
\end{figure}

The formation energy of a bare NiFe alloy substrate $E_{\rm form}$ is as
defined as 
\begin{equation}
E_{\rm form}^{\rm surf} = \frac{- E_{\rm NiFe}^{(9)}+E_{\rm NiFe}^{(7)}}{N} + \alpha \mu_{\rm Ni} + (1-\alpha) \mu_{\rm Fe},
\label{eqn:formationenergy}
\end{equation}
where $E_{\rm NiFe}^{(7)}$ is the total energy of seven layers of Ni$_\beta$Fe$_{1-\beta}$, $N(=8)$ is the number of metal atoms in the surface layers, and $\mu_{\rm Ni}$ ($\mu_{\rm Fe}$) is the chemical potential of the Ni (Fe) atom, which is taken from an fcc (bcc) bulk. Equation~(\ref{eqn:formationenergy}) indicates that the surface with large formation energy is preferable. The atoms indicated by 1 and 2 (1) in Fig.~\ref{fig:Adsorption sites}(a) are Fe in the case of $\alpha$=0.50 ($\alpha$=0.75). Figure~\ref{fig:Total energy of NiFe} shows the formation energy of the bare NiFe alloy substrate. The formation energies of the surface layer of $\alpha=0.75$ are the largest and the substrates with the Ni-rich surface layer ($\alpha > 0.50$) are more preferable than those with the Fe-rich surface layer for all the binary compounds of the bulk layer. This can be explained by the formation energy of the bulk.
\begin{equation}
E_{\rm form}^{\rm bulk} = - E_{\rm NiFe}^{\rm bulk}/(N_{\rm Ni}+ N_{\rm Fe}) + \beta \mu_{\rm Ni} + (1-\beta) \mu_{\rm Fe},
\label{eqn:formationenergyalloy}
\end{equation}
where $E_{\rm NiFe}^{\rm bulk}$ is the total energy of a unit cell of Ni$_\beta$Fe$_{1-\beta}$ bulk and $N_{\rm Ni}$ ($N_{\rm Fe}$) is the number of Ni (Fe) atoms in the unit cell. It is found that the formation energy of the $L1_2$-Ni$_3$Fe bulk is the largest among fcc-Ni (0.000 eV/atom), $L1_2$-Ni$_3$Fe (0.096 eV/atom), $L1_0$-NiFe (0.084 eV/atom), and bcc-Fe (0.000 eV/atom) bulks, which is consistent with the composition ratio of the surface layers.

\subsection{Adsorption energy of graphene on NiFe alloy substrate}
\label{subsec:Adsorption energy of graphene on NiFe alloy substrate}
\begin{figure}[htbp]
\includegraphics{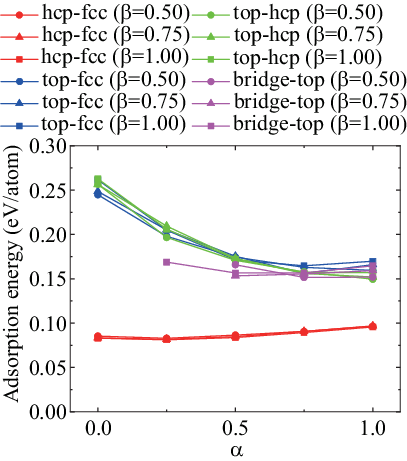}
\caption{Adsorption energy of graphene on NiFe alloy substrate.}
\label{fig:Adsorption energy of graphene/NiFe}
\end{figure}

The adsorption energy of graphene on the NiFe alloy substrate $E_{\rm ads}$ is defined as
\begin{equation}
    E_{\rm ads} = \frac{- E_{\rm Gr/NiFe} + E_{\rm NiFe} + 2 E_{\rm Gr}}{N_C}, 
\label{eqn:ads}
\end{equation}
where $E_{\rm NiFe}$ is the total energy of nine layers of the NiFe alloy substrate, $E_{\rm Gr}$ is the total energy of the graphene monolayer, $E_{\rm Gr/NiFe}$ is the total energy of the NiFe alloy substrate with the graphene monolayer at both ends, and $N_C(=16)$ is the number of C atoms in the supercell. According to Eq.~(\ref{eqn:ads}), the large adsorption energy indicates a stable structure. The adsorption energy is shown in Fig.~\ref{fig:Adsorption energy of graphene/NiFe}. In the case of the Fe-rich surface, the graphene adsorbed at the bridge-top site is relaxed up to the top-fcc or top-hcp site. The adsorption energy is the largest when graphene is adsorbed on the Fe-rich surface at the top-hcp site. On the other hand, in the case of the Ni-rich surface, the graphene adsorbed at the top-fcc or bridge-top site is stable and the difference in the adsorption energy among top-hcp, top-fcc and bridge-top sites is small. The distance $d$ between the atoms in the surface layer and graphene is $ca$ 2.1 \AA. Regardless of the adsorption sites and binary compounds of the bulk layer, the interfaces with the surface layer of $\alpha$=0.00 are the most stable. It is noteworthy that these results do not correspond to the stability of the bare NiFe alloy substrate with respect to the composition ratio of the surface layer.

\subsection{PDOS of interface}
\label{subsec:Partial density of states of interface}
\begin{figure}
\includegraphics{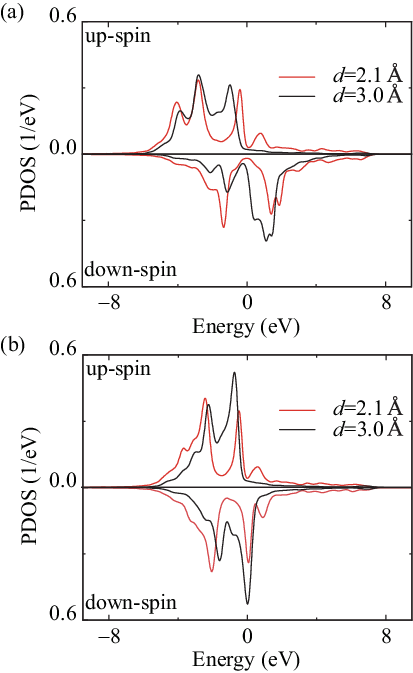}
\caption{PDOSs of d$_{z^2}$ orbitals of (a) surface Fe atom and (b) surface Ni atom where graphene is adsorbed on top-hcp site.}
\label{fig:pdos}
\end{figure}

The increase in adsorption energy with respect to $\alpha$ shown in Fig.~\ref{fig:Adsorption energy of graphene/NiFe} can be explained by the PDOS. The PDOSs of the interface where graphene is adsorbed at the top-hcp site of the Ni substrate ($\beta$=1.00) with $\alpha$=0.00 and 1.00 are plotted in Fig.~\ref{fig:pdos}. The positions of C atoms in graphene (metal atoms in the surface layers) are fixed at those of isolated graphene (ideal surface). To compare the electronic structure of the surface layer before and after the adsorption, the PDOSs with the large [$d$=$d_{\rm desorp}$(=3.0 \AA)] and small interlayer distances [$d$=$d_{\rm adsorp}$(=2.1 \AA)] are plotted. When the interlayer distance is large, the difference in adsorption energy between the substrates with the Fe surface layer (0.251 eV/C atom) and Ni surface layer (0.234 eV/C atom) is small. Lu {\it et al.}\cite{ApplPhysRev_8_031307} reported that the interface with over $d$=3.0 \AA \hspace{1mm} shows physisorption, in which only the van der Waals interaction exists between the atoms in the surface layer and graphene, and the atoms with the interlayer distance of $d$$\approx$2.1 \AA \hspace{1mm} are chemisorbed. In Fig.~\ref{fig:pdos}, the occupation number of the $d_{z^2}$ down-spin orbitals of the surface Fe atoms is small before the adsorption, whereas the occupation number increases after the adsorption, indicating that there are hybridizations between the $d_{z^2}$ orbitals of the Fe atoms in the surface layer and the $p_z$ orbitals of the C atoms in graphene. On the other hand, the $d_{z^2}$ down-spin orbitals of the surface Ni atoms are occupied before the adsorption. Thus, the energy gain of the adsorption on the Ni surface layer is negligible. The hybridization of the $d_{z^2}$ and $p_z$ orbitals is consistent with the result that the graphene at the bridge-top site moves to top-hcp or top-fcc site on Fe-rich surface. Our result agrees with the previous reports that the bridge-top site is the most stable for a Ni substrate,\cite{JPhysChemC_116_007360} while the atop site is preferable when an Fe layer is inserted between the substrate and graphene.\cite{PhysRevB_88_165410}

\begin{figure}[htbp]
\includegraphics{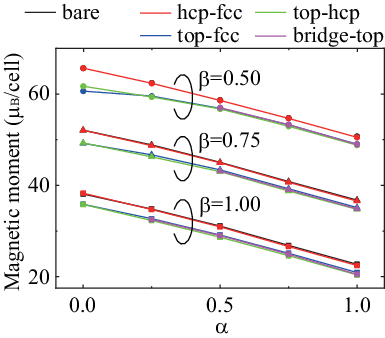}
\caption{Magnetic moments of bare substrate and graphene adsorbed on substrate.}
\label{fig:magnetic moments}
\end{figure}

\begin{figure}
\includegraphics{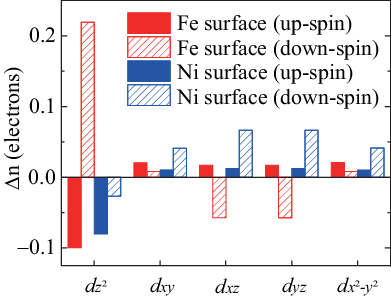}
\caption{Difference in occupation between $d$=2.1 \AA \hspace{1mm} and $d$=3.0 \AA.}
\label{fig:diffcharge}
\end{figure}

The magnetic moments decrease when graphene is adsorbed at the top-hcp and top-fcc sites as shown in Fig.~\ref{fig:magnetic moments}. We then introduce the occupation of the atomic orbitals:
\begin{equation}
n^a_{j,s}(d)= \int_{-\infty}^{E_F} \rho^a_{j,s}(d) dE,
\end{equation}
where $a$ is the index of atoms, $j(= d_{z^2}, d_{xy}, d_{xz}, d_{yz}, d_{x^2-y^2})$ is the index of atomic orbitals, $s(=\uparrow$ or $\downarrow)$ is the index of spins, $\rho^a_{j,s}(d)$ is the PDOS with interlayer distance $d$, and $E_F$ is the Fermi energy. Figure~\ref{fig:diffcharge} shows the difference in the occupations before and after the adsorption.
\begin{equation}
\Delta n^a_{j,s} = n^a_{j,s} (d_{adsorp})- n^a_{j,s} (d_{desorp})
\end{equation}
The occupations of the $d_{z^2}$ down-spin orbitals of the surface Fe atoms and the $d_{xy}$, $d_{yz}$, $d_{zx}$, and $d_{x^2-y^2}$ down-spin orbitals of the surface Ni atoms increase after the adsorption, whereas those of the up-spin orbitals decrease as shown in Fig.~\ref{fig:diffcharge}, resulting in the decrease in the magnetic moment of the interface. When graphene is adsorbed on the hcp-fcc site, the adsorption energy and the change in magnetic moment are significantly smaller than those in the other cases because no hybridization occurs between the $d$ and $p_z$ orbitals. The Fe atoms in the surface layer play important role for the adsorption of graphene on a NiFe alloy substrate. The Ni rich surface is formed when the FM metal is solely grown, and the graphene is transferred on the Ni rich surface. However, when the FM metal is evaporated on graphene, Fe atoms are preferentially adsorbed on graphene. Our results indicate that the fabrication process affects the composition ratio of the interface of graphene/NiFe alloy substrate.

\section{Conclusion}
\label{sec:Conclusion}
We have investigated the adsorption energy of graphene on the NiFe alloy substrate. Although the NiFe alloy substrate with a Ni-rich surface layer is stable, the adsorption energy on the NiFe alloy substrate with the Fe surface layer is smaller than that with the Ni surface layer. The large adsorption energy on the Fe surface layer is interpreted to be due to the hybridization of the $d_{z^2}$ orbitals of metal atoms and the $p_z$ orbitals of C atoms. The occupation number of the $d_{z^2}$ down-spin orbitals of surface Fe atoms increases so as to form the bonding state with the $p_z$ orbitals of C atoms after the adsorption. On the other hand, in the case of the substrate with the Ni surface layer, the $d_{z^2}$ down-spin orbitals of surface Ni atoms are almost occupied before the adsorption. The interface atomic structure where the C atom is adsorbed on the top site of the surface layer is more preferable than that on the fcc or hcp hollow site, and the magnetic moment of the surface layer decreases owing to the hybridization. The occupation numbers of the $d$ down-spin orbitals of the atoms in the surface layer increase after the adsorption, whereas those of the up-spin orbitals decrease, resulting in the decrease of magnetic moment. When the C atoms are adsorbed on the fcc-hpc sites, no hybridizations occur and the magnetic moment of the surface does not decrease. For the fabrication of practical MTJ devices, two processes are proposed. One is the transfer process, where exfoliated 2D materials are transferred to the NiFe alloy electrode from other materials. The other process is the evaporation of the NiFe alloy onto the suspended graphene flake. The Ni-rich surface layer is formed in the former process, whereas the Fe layers are preferentially formed on graphene in the latter process, indicating that the interface atomic structures change depending on the fabrication process. The work in progress entails the investigation of the relationship between the spin-dependent transport property and the electronic structure of the interface by first-principles transport-property calculations.\cite{KikujiHirose2005}

\acknowledgments
This work was supported as part of the JSPS KAKENHI (JP22H05463, JP24K06922, JP24H01196, JP24K01346), JST CREST (JPMJCR22B4), and JSPS Core-to-Core Program (JPJSCCA20230005). The numerical calculations were carried out using the computer facilities of the Institute for Solid State Physics at The University of Tokyo, the Center for Computational Sciences at University of Tsukuba, and the supercomputer Fugaku provided by the RIKEN Center for Computational Science (Project ID: hp230175, hp240178, hp250193).

\bibliography{main}
\bibliographystyle{apsrev4-1}
\end{document}